\documentclass[11pt]{article}
\usepackage{latexsym,array,theorem,mathrsfs,bm,float}
\usepackage{psfrag}
\usepackage{amsfonts,amsmath,amssymb,latexsym,array,afterpage,
theorem,mathrsfs,bm,float,epsfig,color,graphicx,tabularx,
here,multirow,mediabb}
\usepackage[dvipdfmx]{hyperref}
\usepackage{fancybox}
  \makeatletter
    
    \@addtoreset{equation}{section}
  \makeatother

\usepackage[top=30mm,bottom=35mm,left=25mm,right=25mm]{geometry}

\newcommand{\bea}{\begin{eqnarray}}
\newcommand{\ena}{\end{eqnarray}}
\newcommand{\beann}{\begin{eqnarray*}}
\newcommand{\enann}{\end{eqnarray*}}

\newcommand{\ma}[1]{\mbox{$\mathcal{#1}$}}

\newcommand{\ti}{\tilde}

\newcommand{\calhR}[1]{\raisebox{2ex}{\tiny ({\em h})}\hspace{-0.8em}{\ma R}}
\newcommand{\pd}{\partial}
\newcommand{\BS}{\boldsymbol}
\newcommand{\MC}{\mathcal}

\newcommand{\MB}{\mathbb}

\newcommand{\p}{\prime}
\newcommand{\DS}{\displaystyle}

\newif\iffigure
\figurefalse
\figuretrue

\begin{document}

\title{\LARGE{\bf{
Thermodynamics of the 3-dimensional Einstein-Maxwell system}}
}

\author{
\\
Shoichiro Miyashita\thanks{e-mail address : s-miyashita"at"gms.ndhu.edu.tw} 
\\ \\ 
{\it Department of Physics, National Dong Hwa University, Hualien, Taiwan, R.O.C.} 
\\ \\
}

\date{~}

%\thisfancyput(13.cm,1cm){{\fbox{WUCG-22-??}}}

\maketitle
\begin{abstract}
Recently, I studied the thermodynamical properties of the Einstein-Maxwell system with a box boundary in 4-dimensions \cite{Miyashitanew}. In this paper, I investigate those in 3-dimensions using the zero-loop saddle-point approximation and focusing only on a simple topology sector as usual. Similar to the 4-dimensional case, the system is thermodynamically well-behaved when $\Lambda<0$ (due to the contribution of the ``bag of gold'' saddles). However, when $\Lambda=0$, a crucial difference to the 4-dimensional case appears, i.e. the 3-dimensional system turns out to be thermodynamically unstable, while the 4-dimensional one is thermodynamically stable. This may offer two options for how we think about the thermodynamics of 3-dimensional gravity with $\Lambda=0$. One is that the zero-loop approximation or restricting the simple topology sector is not sufficient for 3-dimensions with $\Lambda=0$. The other is that 3-dimensional gravity is really thermodynamically unstable when $\Lambda=0$.
\end{abstract}

\clearpage

\clearpage

%======================================%
%<<<<<<<<<<<< SECTION I  >>>>>>>>>>>>>>%
%======================================%
%%%%%%%%%%%%%%%%%%%%%%%%%%%%%%%%%%%%%%%%%%%
%%%%%%%%%%%%%%%%%%%%%%%%%%%%%%%%%%%%%%%%%%%
%%%%%%%%%%%%%%%%%%%%%%%%%%%%%%%%%%%%%%%%%%%
%%%%%%%%%%%%%%%%%%%%%%%%%%%%%%%%%%%%%%%%%%%
\section{Introduction}
%%%%%%%%%%%%%%%%%%%%%%%%%%%%%%%%%%%%%%%%%%%%%%%%%%%
%%%%%%%%%%%%%%%%%%%%%%%%%%%%%%%%%%%%%%%%%%%%%%%%%%%
%%%%%%%%%%%%%%%%%%%%%%%%%%%%%%%%%%%%%%%%%%%%%%%%%%%

Although there are various problems, such as the non-renormalizability problem, the integration contour problem \cite{GibbonsHawkingPerry}, gravitational path integral may be useful for extracting the information or challenging the paradoxes of quantum gravity. 
%%%%%% footnote %%%%%%%%%%
\footnote{
One of the most exciting recent developments is the explanation of the Page curve \cite{Page} by using the Euclidean gravitational path integral \cite{PeningtonShenkerStanfordYang, AlmheiriHartmanMaldacenaShaghoulianTajdini, FaulknerLewkowyczMaldacena, LewkowyczMaldacena}. (More precisely, they are the derivation of the island conjecture \cite{AlmheiriMahajanMaldacenaZhao}.) For the details, see the nice review \cite{AlmheiriHartmanMaldacenaShaghoulianTajdini2}.
}
%%%%%%%%%%%%%%%%%%%%%%%

The Euclidean path integral representation of the gravitational partition function is the simplest one \cite{GibbonsHawking}. Formally, its definition is completely parallel to that of quantum field theory except that the external fields are defined on the boundary as the boundary conditions of the bulk fields and we sum over all configurations (and topologies) in the bulk \cite{GibbonsHawking, HawkingPage, York}. In practice, we use the saddle-point approximation and consider only the simplest topologies $S^{1} \times \MB{R}^{d-1}$ and $\MB{R}^{2} \times S^{d-2}$ (for $d$-dimensional spacetime with a spherical boundary), assuming that they give the dominant contributions. It turns out that there exists the Hawking-Page phase structure for the systems with the asymptotically AdS boundary condition \cite{HawkingPage} and with the box boundary condition \cite{York} for pure gravity theory with $\Lambda \leq 0$. Despite of the ignorance of the other topology sectors, this non-trivial phase structure is widely accepted as true and it is sufficient to consider the simplest topologies to describe the thermal equilibrium of quantum gravity, at least for an appropriate parameter region. 
%%%%%%%%%%% footnote %%%%%%%%%%%%%%
\footnote{
$(-\Lambda)^{\frac{d-2}{2}} G \gg 1$ for the asymptotically AdS boundary condition and  $r_{b}^{d-2}/G \gg 1$ for the box boundary condition (where $r_{b}$ is the radius of the boundary sphere).
}
%%%%%%%%%%%%%%%%%%%%%%%%%%%%%%%%%
 (For 2-dimensional gravity theories, however, see an interesting discussion \cite{EngelhardtFischettiMaloney} on the importance of the higher topology sectors.) So the task boils down to just finding the solutions $\BS{g}_{i}$ in the simple topology sectors, evaluating the action of the solutions $I^{E}[\BS{g}_{i}]$, and summing over $e^{-I^{E}[\BS{g}_{i}]}$s (or just picking the dominant one). 
 
 Even in this simplified task, it has caused problems in the literature. One is the integration contour problem \cite{GibbonsHawkingPerry}. Another, which I would like to treat as a problem, is the problem of the ``bag of gold (BG)'' saddle \cite{Miyashita, Miyashitanew}. This problem arises when we consider the box boundary conditions. For both the AdS boundary condition and the box boundary condition, (Euclidean) black hole (BH) geometries become the saddle points in the $\MB{R}^{2} \times S^{d-2}$ topology sector. Its Lorentzian interpretations may be ``a BH in AdS'' and ``a BH in a box'',  respectively. However, for some systems with the box boundary condition, there exists another type of saddles in the topology sector, which I called the``bag of gold (BG)'' saddle \cite{Miyashita}. The difference to the BH saddle is that the area of the bolt is larger than that of the boundary sphere. Part of the problem is that, since their Lorentzian picture is that a horizon is located on the ``outside'' of a box, people have implicitly or explicitly excluded the contributions of BG saddles by ``physical'' considerations.
 %%%%%%%%%%%% footnote %%%%%%%%%%%%
\footnote{
The real problem of the BG saddle is that whether all BGs contribute to the partition function or not. As I will explain shortly, it seems that BGs should contribute to the partition function of the Einstein-Maxwell system with $\Lambda\leq0$ from the thermodynamic point of view \cite{Miyashitanew}. However, BGs appearing when $\Lambda>0$ (for both pure gravity and the Einstein-Maxwell) make the system thermodynamically unstable \cite{Miyashita, DraperFarkas, BanihashemiJacobson, Miyashitanew}. 
} 
 %%%%%%%%%%%%%%%%%%%%%%%%%%%%%%%%
An example where the ignorance of BG saddles causes a problem is the Einstein-Maxwell system \cite{BasuKrishnanSubramanian}. In the paper, they showed that, without the contributions of the BG saddles, the free energy behaves somewhat peculiarly, i.e. it becomes a discontinuous function. Recently, I re-examined the thermodynamical properties of the Einstein-Maxwell system in 4-dimensions with the box boundary condition \cite{Miyashitanew} and found that there exist BG saddles and that they contribute to the partition function. As a result, the peculiar behavior of the free energy observed in \cite{BasuKrishnanSubramanian} was resolved and it is shown that the system has a well-defined thermodynamic description for $\Lambda \leq 0$. 

In this paper, I investigate the thermodynamical properties of the Einstein-Maxwell system in 3-dimensions with the box boundary condition. Previously, the case of $\Lambda<0$ was studied by Huang and Tao \cite{HuangTao}, but they did not consider the contribution of BG saddles. So I will start with revisiting the case of $\Lambda<0$ and see how their results are modified by including the BG saddle contributions and how they differ from the 4-dimensional case. Although the analysis is completely parallel to that of 4-dimensions, related to the fact that we do not have BH solutions for $\Lambda\geq0$ in 3-dimensions \cite{Ida, BanadosTeitelboimZanelli}, the thermodynamical properties are qualitatively different from the 4-(or higher )dimensional Einstein-Maxwell system when $\Lambda$ is negative but close to zero or when $\Lambda \geq 0$. A short summary of the result is;
\begin{itemize}
\item When $\Lambda<0$, the system is well-behaved and thermodynamically stable due to the contribution of the BG saddles. Without them, the free energy shows a strange behavior similar to the 4-dimensional case \cite{BasuKrishnanSubramanian}.

\item When $\Lambda<0$ and $-\Lambda r_{b}^2$ is sufficiently large, the phase diagram is similar to the 4-dimensional case \cite{Miyashitanew}. (Fig. \ref{4})

\item When $\Lambda<0$ and $-\Lambda r_{b}^2$ is small, the fraction of the BH phase in the phase diagram shrinks and vanishes when $\Lambda =0$. This is in contrast to the 4-dimensional case, where the BH phase still exists when $\Lambda=0$. (Fig. \ref{4})

\item Therefore, when $\Lambda=0$, only BG saddles
%%%%%%%%%% \footnote %%%%%%%%%%%%%
\footnote{
Precisely.\, BG and Bertotti-Robinson (BR) saddles.
}
%%%%%%%%%%%%%%%%%%%%%%%%%%%%%%%%
 exist in the $\MB{R}^2 \times S^{1}$ sector. (Fig. \ref{5}) But they are not thermodynamically stable. So the system is not thermodynamically stable either, at least at zero-loop order. This is a striking difference to the 4-dimensional case.
\end{itemize}

As I noted above, we usually believe that (i) the (zero-loop) saddle-point approximation, and (ii) the concentration on the simple topology sector, are sufficient to describe the thermodynamics of quantum gravity. This may be true for higher dimensions. However, the last result indicates that, in 3-dimensions with $\Lambda=0$ (at least for the Einstein-Maxwell system), (i) and (ii) are not sufficient. We may have to consider  one-loop corrections or the contribution from the complicated topologies. 
%%%%%%%%%% \footnote %%%%%%%%%%%%%
\footnote{
Or believing that (i) and (ii) are sufficient, conclude that the system is truly thermodynamically unstable.
}
%%%%%%%%%%%%%%%%%%%%%%%%%%%%%%%%

The organization of this paper is as follows: 
In Section 2, I study the thermodynamical properties of the Einstein-Maxwell system with box boundary condition in 3-dimensions when $\Lambda<0$. In subsection 2.1, I review the analysis of Huang and Tao \cite{HuangTao}, in which they investigated the detailed properties of the empty saddles and the BH saddles of the system. Although the analysis on these saddles was correct, their two additional assumptions lead to the incorrect results on the phase structure of the system. In subsection 2.2, I point out that the Bertotti-Robinson (BR) saddles \cite{Bertotti, Robinson, Clement, ClementFabbri} and the BG saddles were missing in their analysis and investigate their properties. I then show the complete phase structure of the system. In Section 3, I point out a qualitative difference on the behavior of phase structure between the 3-dimensional and the 4-dimensional system when $\Lambda r_{b}^2$ is close to zero. In Section 4, I investigate the thermodynamical properties of the Einstein-Maxwell system with box boundary condition in 3-dimensions when $\Lambda=0$. Throughout this paper, I concentrate only on the grand canonical ensembles.

%======================================%
%<<<<<<<<<<<< SECTION II  >>>>>>>>>>>>>>%
%======================================%
%%%%%%%%%%%%%%%%%%%%%%%%%%%%%%%%%%%%%%%%%%%
%%%%%%%%%%%%%%%%%%%%%%%%%%%%%%%%%%%%%%%%%%%
%%%%%%%%%%%%%%%%%%%%%%%%%%%%%%%%%%%%%%%%%%%
%%%%%%%%%%%%%%%%%%%%%%%%%%%%%%%%%%%%%%%%%%%
\section{$\Lambda < 0$ case}
%%%%%%%%%%%%%%%%%%%%%%%%%%%%%%%%%%%%%%%%%%%%%%%%%%%
%%%%%%%%%%%%%%%%%%%%%%%%%%%%%%%%%%%%%%%%%%%%%%%%%%%
%%%%%%%%%%%%%%%%%%%%%%%%%%%%%%%%%%%%%%%%%%%%%%%%%%%
In this section, I study the thermodynamical properties of the 3-dimensional Einstein-Maxwell system with box boundary and $\Lambda <0$. In 3-dimensions, it is often said that there is no BH when $\Lambda \geq0$, and only when $\Lambda <0$, BTZ BHs can exist. Therefore it can be a good starting point for studying thermodynamics in 3-dimensions. And indeed, in \cite{HuangTao}, they considered this case and investigated thermodynamic properties. In subsection 2.1, the basic setups and the properties of empty saddles and BH saddles are reviewed, which are systematically studied in \cite{HuangTao}. For empty saddles and BH saddles, their results are correct. But for other states, they made wrong statements. As a result, some part of their results on the phase diagram are incorrect. In subsection 2.2, I point out that  BR saddles and BG saddles were missing in their investigation and examine their properties. Finally, I show the complete phase diagram.

%%%%%%%%%%%%%%%%%%%%%%%%%%
\subsection{empty saddles and BH saddles}
%%%%%%%%%%%%%%%%%%%%%%%%%%
The Euclidean action for grand canonical ensembles \cite{BradenBrownWhitingYork, HawkingRoss} is
%%%%%%%%%%%%%% footnote %%%%%%%%%%%%%%%%
\footnote{
The last term becomes the standard counterterm for asymptotically AdS spacetime \cite{BalasubramanianKraus, EmparanJohnsonMyers} when $\Lambda<0$, i.e. $\frac{1}{8\pi G} \int_{\pd\MC{M}} d^2 y \sqrt{\gamma}  \sqrt{|\Lambda|}=\frac{1}{8\pi G} \int_{\pd\MC{M}} d^2 y \sqrt{\gamma} \frac{1}{l_{AdS}}$ where $l_{AdS}$ is the AdS radius. I used $\sqrt{|\Lambda|}$ instead of $1/l_{AdS}$ because it can be used for any value of $\Lambda$. Note that all quantities in this paper do not have divergences because the boundary sphere is of finite volume. So the last term is not necessary. It is just a constant shift of free energy and energy and does not affect other thermodynamical quantities.  However, only if there exists the counterterm, we can take a limit $r_{b}/l_{AdS}\to \infty$ with suitable rescaling of thermodynamical quantities and recover the thermodynamics of asymptotically AdS boundary condition. (See subsection 4.1 of \cite{Miyashitanew}.) 
}
%%%%%%%%%%%%%%%%%%%%%%%%%%%%%%%%%%%%%%
\bea
I^{E}[\BS{g}, \BS{A}]= \frac{-1}{16\pi G} \int_{\MC{M}} d^3 x \sqrt{g} (\MC{R}- 2\Lambda) + \frac{1}{16 \pi}\int_{\MC{M}} d^3 x \sqrt{g} F_{\mu\nu} F^{\mu\nu} \hspace{2.5cm} \notag \\
+ \frac{-1}{8\pi G} \int_{\pd\MC{M}} d^2 y \sqrt{\gamma} \Theta + \frac{1}{8\pi G} \int_{\pd\MC{M}} d^2 y \sqrt{\gamma}  \sqrt{|\Lambda|} \label{action}
\ena
I focus only on the metric and the gauge field of the form
\bea
ds^2 = g_{\mu\nu}dx^{\mu}dx^{\nu}= f(r)dt^2 + \frac{1}{f(r)}dr^2 + r^2 d\phi^2 \\
A_{\mu}dx^{\mu}= A_{t}(r)dt
\ena
and assume that they are the dominant contributions to the Euclidean path integral representation of the grand canonical partition function $Z(\beta, \mu, r_{b})$ of gravity. And I also assume that they are purely ``Euclidean'', meaning that the metric is the real Euclidean metric and the gauge field is purely imaginary. 
%%%%%%%%%%%% footnote %%%%%%%%%%%
\footnote{
In order to keep the electric charge real in the Euclidean picture, the boundary value of $A_{t}$ must be imaginary on the boundary. Therefore the simplest Euclidean solutions may be of real Euclidean metric and purely imaginary gauge field, since these correspond to the well-known Lorentzian solutions after the Wick rotation. These may be the analog of purely Euclidean solutions for the pure gravity case. For a little more detail, see the footnote 11 of $\cite{Witten}$, where the similar problem of complex configurations for the rotating grand canonical ensemble was discussed.
}
%%%%%%%%%%%%%%%%%%%%%%%%%%%%%%%
Here, $r$ is the areal radius and the boundary at $r=r_{b}$. Within this class of metrics and gauge fields, there are at least two types of solutions. One is the empty saddle;
\bea
{\rm Empty ~ saddle:}~~  f(r) =  1-\Lambda r^2 \hspace{3.7cm} \\
A_{t}(r)= -i\ti{\mu}  \hspace{4.2cm}
\ena
The other is the black hole (BH) saddle.
%%%%%%%%%%%% footnote %%%%%%%%%%%
\footnote{
Although in the standard parametrization of solutions I should say that there is another type of saddles
\bea
f(r) =  -\Lambda r^2   \notag \\
A_{t}(r)= -i\ti{\mu}   \notag 
\ena
which corresponds to the Poincare patch of AdS, we can obtain this from $r_{H} \to 0$ limit of BH saddles in the parametrization in the grand canonical ensembles. As I will explain shortly, the parameter $\ti{Q}$ in (\ref{BHf}) and (\ref{BHA}) can be written as
\beann
\ti{Q}=  \frac{\mu}{2} \sqrt{ \frac{ - \Lambda (r_{b}^2 - r_{H}^2) }{ \left(2 G \mu^2 +  \log \frac{r_{b}}{r_{H}} \right) \log \frac{r_{b}}{r_{H}} } }
\enann
With this expression, we can show that
\beann
({\rm the ~ last ~ term ~ in ~ (\ref{BHf})}) \to 0 \\
A_{t} \to -i \sqrt{-\Lambda} r_{b} \mu = -i \ti{\mu}
\enann
In addition, unlike the BH saddles, this saddle can have any temperature. However, the free energy of this saddle is always larger than that of the empty saddle (i.e. $F_{Poincare}=0, ~ F_{empty}= -\frac{\sqrt{1-\Lambda r_{b}^2}}{4G} + \frac{\sqrt{|\Lambda|}}{4G} r_{b}  <0 $.) So I will ignore this saddle in this paper.
}
%%%%%%%%%%%%%%%%%%%%%%%%%%%%%%%
\bea
{\rm BH ~ saddle:}  ~~ f(r) =  \Lambda r_{H}^2 -\Lambda r^2 -  8G\ti{Q}^2 \log \frac{r}{r_{H}} \label{BHf} \\
A_{t}(r) = -2 i \ti{Q} \log \frac{r}{r_{H}} \hspace{2.1cm} \label{BHA}
\ena
where $\ti{\mu}, \ti{Q}$ and $r_{H}\in (0, r_{b})$ are parameters that depend on $\beta, \mu, r_{b}$ through the boundary condition of the path integral for the grand canonical partition function $Z(\beta, \mu, r_{b})$ \cite{BradenBrownWhitingYork};
\bea
{\rm \underline{boundary ~ condition}} \hspace{11.2cm} \notag \\
{\rm Empty ~ saddle:} ~~ \frac{\ti{\mu}}{\sqrt{f(r_{b})}} = \mu    \hspace{7.6cm} \label{bdymu} \\
{\rm BH ~ saddle:} ~~ \frac{4 \pi}{f^{\p}(r_{H})}\sqrt{f(r_{b})}= \beta, ~~ \frac{2\ti{Q}}{\sqrt{f(r_{b})}} \log \frac{r_{b}}{r_{H}} =\mu \hspace{2.85cm} \label{bdyT}
\ena
In the following, the function $f$ will be used only for the BH (and BG) saddles. Substituting these field configurations to the action (\ref{action}), we get 
\bea
{\rm \underline{free ~ energy}} \hspace{12.7cm} \notag \\
{\rm Empty ~ saddle:} ~~ F = -\frac{\sqrt{1-\Lambda r_{b}^2}}{4G} + \frac{\sqrt{|\Lambda|}}{4G} r_{b}   \hspace{6.6cm} \\
{\rm BH ~ saddle:} ~~ F= - \frac{\sqrt{f(r_{b})}}{4G} + \frac{\sqrt{|\Lambda|}}{4G}r_{b} +\frac{\Lambda r_{H}^2}{4G \sqrt{f(r_{b})}} + \frac{\ti{Q}^2}{\sqrt{f(r_{b})}} -\frac{2\ti{Q}^2}{\sqrt{f(r_{b})}} \log \frac{r_{b}}{r_{H}} 
\ena
Again, I want to emphasize that the parameters $r_{H}$ and $\ti{Q}$ are functions of $\beta, \mu, r_{b}$.  For the empty saddle, since it has no  $T$ dependence and $\mu$ dependence, the thermodynamical quantities are given by
\bea
{\rm \underline{thermodynamical ~ quantity}} \hspace{10cm} \notag \\
{\rm Empty ~ saddle:} ~~ E= -\frac{\sqrt{1-\Lambda r_{b}^2}}{4G} + \frac{\sqrt{|\Lambda|}}{4G} r_{b}  \hspace{6.2cm} \\
S=0 \hspace{9.8cm} \\
Q=0 \hspace{9.8cm}
\ena

For the BH saddle, we can obtain them by using the relations $\DS \left(\frac{\pd F}{\pd \beta}\right)_{\mu, r_{b}}=-E, \DS \left(\frac{\pd F}{\pd \mu}\right)_{\beta, r_{b}}=-Q$ and $S=\beta (E-\mu Q - F)$. Alternatively, since we also have finer-grained information about the saddle point geometries, we can obtain them by
\bea
 E= \int d \phi \sqrt{\sigma} u_{i} \tau^{ij} \xi_{j} \\
 Q = \int d \phi \sqrt{\sigma} j^{i}u_{i} 
\ena
where $\tau_{ij}$ and $j^{i}$ are the boundary currents of the metric and the gauge field, defined by $\displaystyle \tau_{ij}\equiv \frac{2}{\sqrt{\gamma}} \frac{\delta I^E}{\delta \gamma^{ij}}, ~~ j^{i}\equiv \frac{i}{\sqrt{\gamma}} \frac{\delta I^E}{\delta A_{i}} $. (The former current is known as the Brown-York tensor \cite{BrownYork}.)
$\xi^{i}$ is the normalized Killing vector of the $U(1)$ isometry on the boundary and $u^i$ is the normal vector of the integration surface. Using these fine-grained definitions and the relation $S=\beta (E-\mu Q - F)$, we get
\bea
{\rm \underline{thermodynamical ~ quantity}} \hspace{10cm} \notag \\
{\rm BH ~ saddle:} ~~ E= -\frac{\sqrt{f(r_{b})}}{4G} + \frac{\sqrt{|\Lambda|}}{4G} r_{b}  \hspace{6.5cm} \\
S= \frac{\pi r_{H}}{2G} \hspace{9.1cm} \\
Q=\ti{Q} \hspace{9.55cm}
\ena

Although they did not insist on this point in \cite{HuangTao}, they showed that there exists a maximum temperature for BHs that can be reached by the limit of the BH horizon approaching the boundary, and that there are no BHs above this temperature. Since they did not show the analytical expression of the temperature, I will derive it here. From the boundary conditions, the temperature can be rewritten as a function of $r_{H}$ (and $\mu$), and its square is 
\bea
T(r_{H})^2= \frac{-\Lambda \left\{ -G \mu^2 (r_{b}^2-r_{H}^2) + \left( 2G \mu^2 + \log \frac{r_{b}}{r_{H}} \right) r_{H}^2 \log \frac{r_{b}}{r_{H}} \right\}^2 }{4\pi^2 r_{H}^2(r_{b}^2 - r_{H}^2) \left(\log \frac{r_{b}}{r_{H}} \right)^3 \left( 2G \mu^2 + \log \frac{r_{b}}{r_{H}} \right)  } \label{T2}
\ena
In order to apply L'H${\rm \hat{o}}$pital's theorem, I define functions $N_{T^2}(r_{H})$ and $D_{T^2}(r_{H})$ by
\bea
N_{T^2}(r_{H}) = ({\rm the ~ Numerator ~ of ~} T(r_{H})^2 ~  (\ref{T2})) \notag \hspace{3.5cm} \\
=-\Lambda \left\{ -G \mu^2 (r_{b}^2-r_{H}^2) + \left( 2G \mu^2 + \log \frac{r_{b}}{r_{H}} \right) r_{H}^2 \log \frac{r_{b}}{r_{H}} \right\}^2  \\
~ \notag \\
D_{T^2}(r_{H}) = ({\rm the ~ Denominator ~ of ~} T(r_{H})^2 ~  (\ref{T2})) \notag \hspace{3.2cm} \\
= 4\pi^2 r_{H}^2(r_{b}^2 - r_{H}^2) \left(\log \frac{r_{b}}{r_{H}} \right)^3 \left( 2G \mu^2 + \log \frac{r_{b}}{r_{H}} \right) \hspace{1.6cm}
\ena
As we can easily check, the fourth derivatives of these functions satisfy
\bea
\lim_{r_{H}\to r_{b}} \frac{N_{T^2}''''(r_{H})}{D_{T^2}''''(r_{H})} = \frac{-\Lambda (1-2G\mu^2)^2 }{16\pi^2 G \mu^2 }
\ena
Using L'H${\rm \hat{o}}$pital's theorem, $\DS \lim_{r_{H}\to r_{b}} T(r_{H})^2 = \frac{-\Lambda (1-2G\mu^2)^2 }{16\pi^2 G \mu^2 } $. Therefore, for fixed $\mu$, $r_{H}\to r_{b}$ limit of temperature is
\bea
T(r_{b})= \frac{\sqrt{-\Lambda}}{4\pi } \frac{1- 2G\mu^2}{\sqrt{G  }\mu} \label{Tend}
\ena 
The existence of a maximum temperature of charged BHs in the Einstein-Maxwell system in a box was pointed out in \cite{BasuKrishnanSubramanian, Miyashitanew} for 4-dimensions, and in \cite{FernandesLemos} for higher dimensions. Similarly, we can derive the maximum charge of BH saddles;
\bea
Q(r_{b})= \frac{r_{b}}{2}\sqrt{\frac{-\Lambda}{G}} 
\ena
Note that $T(r_{b})$ does not depend on $r_{b}$. On the other hand, $Q(r_{b})$ does not depend on $\mu$.

In addition, they have made two claims in \cite{HuangTao} that I would like to claim to be false in the next subsection.\\
~\\
\fbox{Their claim \cite{HuangTao}} (which I will claim to be false)
\begin{itemize}
\item When taking the $r_{H} \to r_{b}$ limit of the BH saddles, it corresponds to the \underline{``M state''.} Its entropy is $S= \frac{\pi r_{b}}{2G}$, its charge is $Q=\frac{r_{b}}{2}\sqrt{\frac{-\Lambda}{G}} $, and its energy is $E= \frac{\sqrt{-\Lambda}}{4G}r_{b} $. \\
And \underline{its temperature $T$ and its chemical potential $\mu$ can be arbitrary values.}
\item There is a critical temperature, above which there are no BH saddles. Above this temperature, \underline{ ``M states'' give the dominant contribution.}
\end{itemize}
 According to their paper, the free energy of the ``M state'' is
\bea
F_{M state}(T, \mu) = \frac{\sqrt{-\Lambda}}{4G}r_{b} - T \frac{\pi r_{b}}{2G} - \mu \frac{r_{b}}{2}\sqrt{\frac{-\Lambda}{G}} \label{FMstate}
\ena  
They expressed this state as a state of ``the black hole merging with the boundary''. An example of the behavior of the free energies of the empty, BH, and M states, and the phase diagram are shown in Fig. \ref{1}. In the next subsection, I will show that the horizon does not merge with the boundary in the $r_{H}\to r_{b}$ limit and can even be larger than the boundary.
%%%%%%%%%%%%%%%%%%
                                                 %
\iffigure
\begin{figure}[h]
\begin{center}
	\includegraphics[width=8.cm]{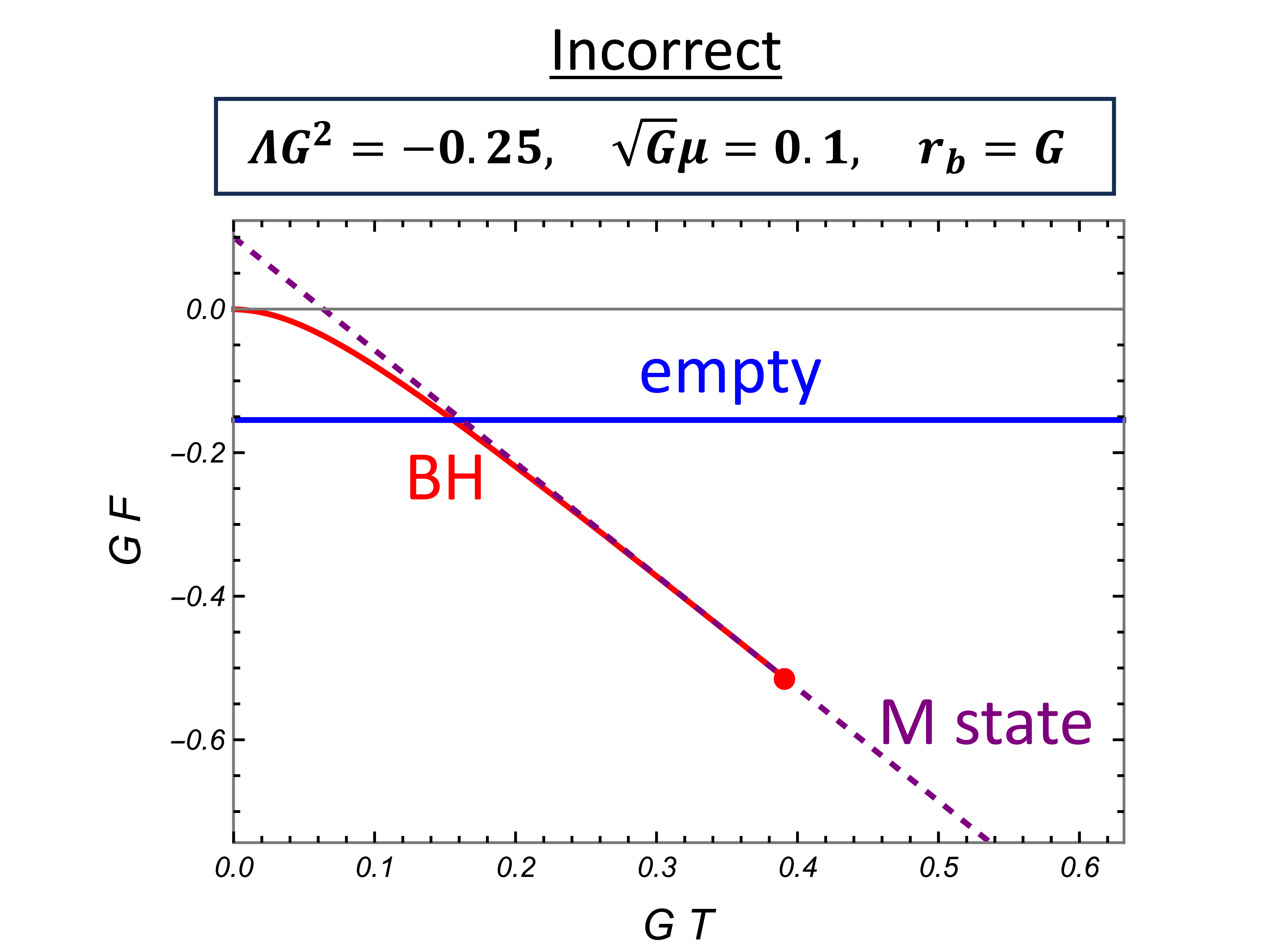} 	\includegraphics[width=8.cm]{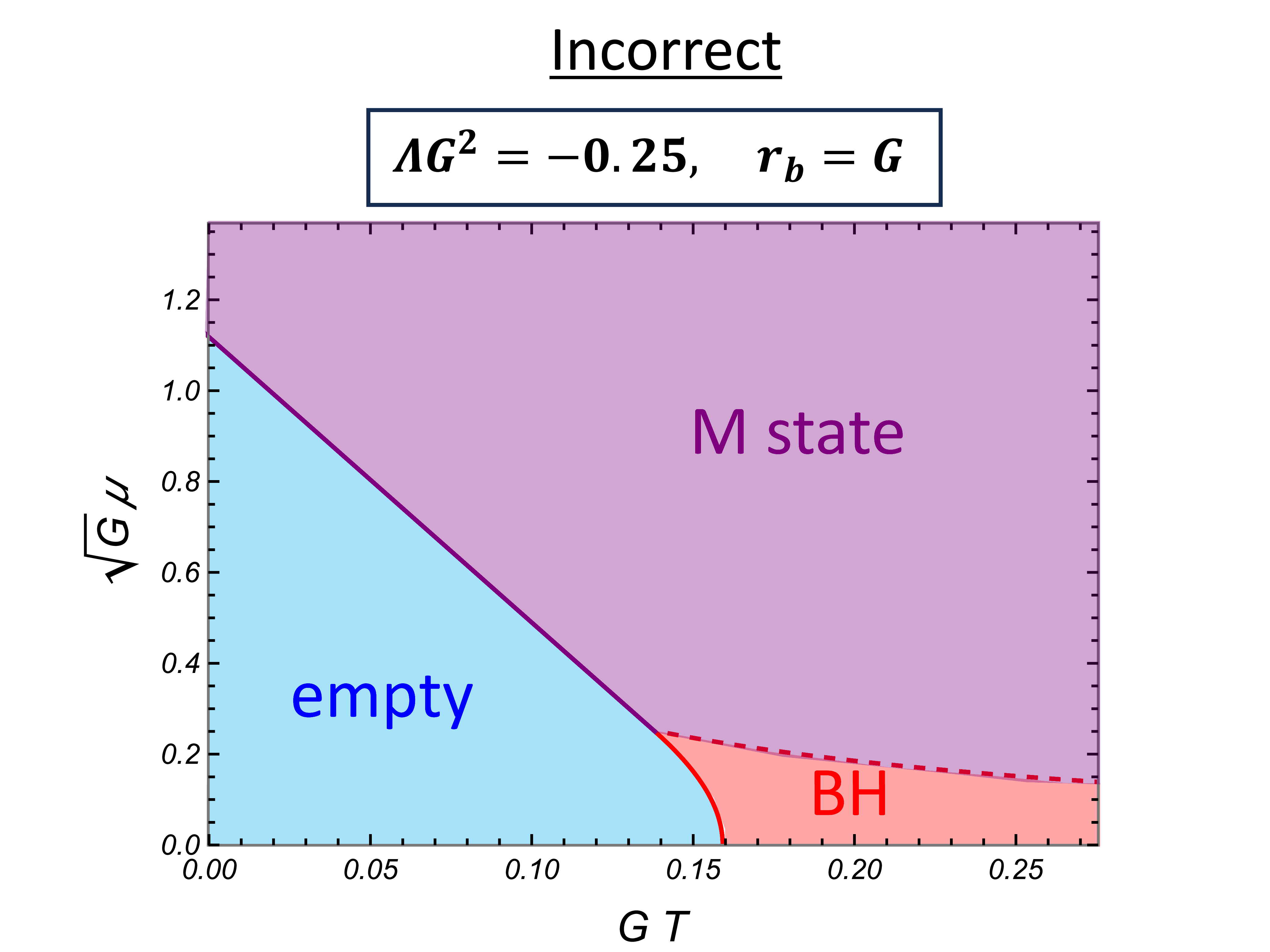} 
	\caption{{\it Incorrect} $F-T$ diagram and phase diagram. (For the empty and BH phases, they are correct.) In \cite{HuangTao}, they claimed that there are two phase transitions; one is the Hawking-Page transition between the empty and BH phases, the other is a second order phase transition between the BH phase and the M state. The transition point is the same as the termination point of the BH branch and it is given by (\ref{Tend}) and marked with the red dot in the left figure. (\ref{Tend}) also represents the boundary between the BH phase and the M states in the right figure. The boundary between the empty phase and the M state is given by the relation $T= \frac{\sqrt{1-\Lambda r_{b}^2 }}{ 2\pi r_{b} } - \frac{\mu}{\pi} \sqrt{-\Lambda G} $. }
\label{1}
\end{center}
\end{figure}
\fi
                                                 %
%%%%%%%%%%%%%%%%%%

%%%%%%%%%%%%%%%%%%%%
%%%%%%%%%%%%%%%%%%%%
%%%%%%%%%%%%%%%%%%%%%%%%%%
\subsection{BR saddles, BG saddles, and thermodynamics}
%%%%%%%%%%%%%%%%%%%%%%%%%%
Instead of their two claims, I would like to claim that the following two are correct;\\
~\\
\fbox{My claim}
\begin{itemize}
\item When taking the $r_{H} \to r_{b}$ limit of the BH saddles, it corresponds to the \underline{Bertotti-Robinson(BR) saddles.} Its entropy is $S= \frac{\pi r_{b}}{2G}$, its charge is $Q=\frac{r_{b}}{2}\sqrt{\frac{-\Lambda}{G}} $, and its energy is $E= \frac{\sqrt{-\Lambda}}{4G}r_{b} $. \\
And \underline{its temperature $T$ and its chemical potential $\mu$ are NOT arbitrary values.}
\item There is a critical temperature, above which there are no BH saddles. Above this temperature, \underline{ ``bag of gold(BG)''  saddles give the dominant contribution.}
\end{itemize}

Firstly, let's show the limit $r_{H} \to r_{b}$ does not correspond to the situation where ``the black hole merges with the cavity'' but to the Bertotti-Robinson (BR) geometry\cite{Bertotti, Robinson}. 
%%%%%%%%%%% footnote %%%%%%%%%%%%%%%
\footnote{
The original Bertotti-Robinson geometry \cite{Bertotti, Robinson} is of 4 dimensions. The 3-dimensional version was obtained in \cite{Clement, ClementFabbri}.
}
%%%%%%%%%%%%%%%%%%%%%%%%%%%%%%%%%%
Define a small parameter $\varepsilon$ by $r_{H}=r_{b}-\varepsilon$. The first and second derivatives of $f(r)$ at $r=r_{H}=r_{b}-\varepsilon$ is
\bea
f'(r_{H}) = \frac{-\Lambda}{G \mu^2} (1-2G \mu^2) \varepsilon + O(\varepsilon^2) \\
f''(r_{H}) = -4\Lambda + O(\varepsilon) \hspace{2.3cm}
\ena 
By defining a new coordinate $r=r_{H} + \varepsilon \ti{r}, t= \frac{\ti{t}}{\varepsilon}$, the metric becomes
\bea
ds^2= \left[ \frac{-\Lambda}{G \mu^2} (1-2G \mu^2) \ti{r} - 2\Lambda \ti{r}^2 +O(\varepsilon) \right] d\ti{t}^2 + \frac{1}{\left[ \frac{-\Lambda}{G \mu^2} (1-2G \mu^2) \ti{r} - 2\Lambda \ti{r}^2 +O(\varepsilon) \right]} d\ti{r}^2 + (r_{b}+ O(\varepsilon))^2 d\phi^2 \notag \\
~ 
\ena
Then, $\varepsilon \to 0$ limit (i.e. near horizon and near extremal limit) leads to
\bea
ds^2= \left[ \frac{-\Lambda}{G \mu^2} (1-2G \mu^2) \ti{r} - 2\Lambda \ti{r}^2 \right] d\ti{t}^2 + \frac{1}{\left[ \frac{-\Lambda}{G \mu^2} (1-2G \mu^2) \ti{r} - 2\Lambda \ti{r}^2 \right]} d\ti{r}^2 + r_{b}^2 d\phi^2 \label{BRmetric} \\
\ti{r}\in[0, 1]
\ena
which is the Euclidean version of the 3-dimensional Bertotti-Robinson (BR) geometry. The gauge field can be obtained in a similar way, i.e,
\bea
\BS{A}= A_{t}(r) dt \hspace{2.5cm} \notag \\
= \left[ \sqrt{\frac{-\Lambda}{G}} \varepsilon \ti{r} + O(\varepsilon^2)  \right] \frac{d\ti{t}}{\varepsilon} \notag \\
\to -i \sqrt{\frac{-\Lambda }{G}} \ti{r} d\ti{t} \hspace{1.6cm}
\ena
As with the standard argument for BHs, the absence of a conical singularity at $\ti{r}=0$ leads to the unique temperature;
\bea
T_{BR} = \frac{\sqrt{-\Lambda}}{4\pi } \frac{1- 2G\mu^2}{\sqrt{G  }\mu} \label{TBR}
\ena
which of course is the same as (\ref{Tend}). We can also check that the parameter $\mu$ is actually the chemical potential of this saddle $\DS \frac{A_{\ti{t}}(\ti{r}=1)}{\sqrt{g_{\ti{t}\ti{t}}}}=\mu$. Free energy and other thermodynamical quantities are given by
 \bea
{\rm \underline{free ~ energy}} \hspace{12.7cm} \notag \\
{\rm BR ~ saddle:} ~~ F= \frac{\sqrt{-\Lambda}}{4G}r_{b} - \frac{r_{b}}{8G\mu}\sqrt{\frac{-\Lambda}{G}} -\frac{r_{b}\mu}{4} \sqrt{ \frac{-\Lambda}{G} } \label{BRfree} \hspace{5cm}
\ena
 \bea
{\rm \underline{thermodynamical ~ quantity}} \hspace{10cm} \notag \\
{\rm BR ~ saddle:} ~~ E= \frac{\sqrt{-\Lambda}}{4G}r_{b} \hspace{9.4cm} \\
 S= \frac{\pi r_{b}}{2G} \hspace{10.05cm} \\
Q= \frac{r_{b}}{2} \sqrt{\frac{-\Lambda}{G}} \hspace{9.25cm}
\ena
%%%%%%%%%%%%%%%%%%
                                                 %
\iffigure
\begin{figure}
\begin{center}
	\includegraphics[width=10.cm]{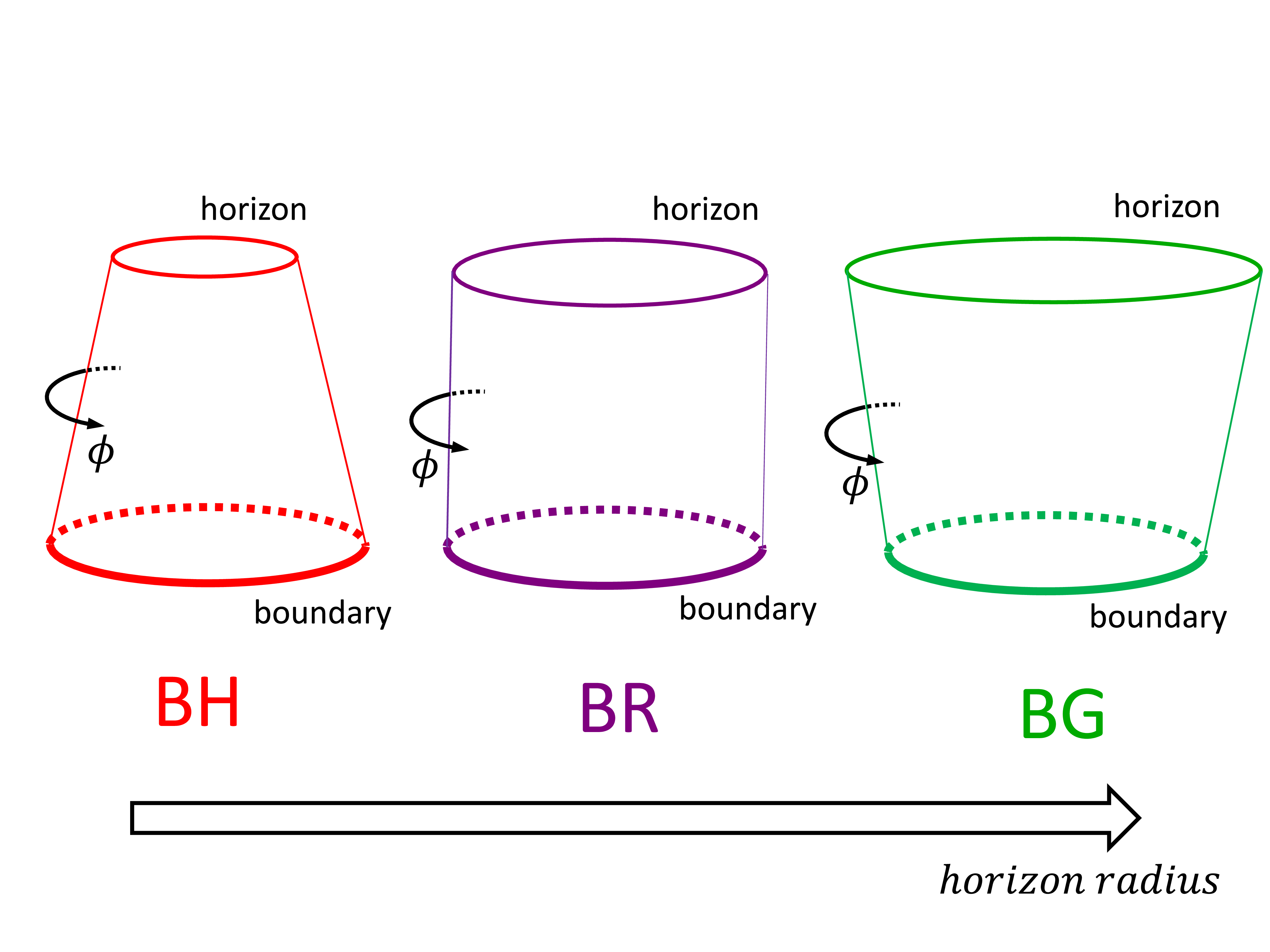} 
	\caption{Spatial section of the BH saddle, BR saddle, and BG saddle. They are characterized by the radius of the horizon (or bolt); whether it is smaller or larger than the radius of the boundary.  }
\label{2}
\end{center}
\end{figure}
\fi
                                                 %
%%%%%%%%%%%%%%%%%%   
So far, we have seen that the $r_{H}\to r_{b}$ limit of BH saddles is not a singular geometry of zero size but a BR saddle, whose circumference of the transverse circle remains constant in the radial direction. Since the limit is not singular, one might wonder whether a further deformation of saddles (i.e. the extension of the parameter range of $r_{H}$ beyond $r_{b}$ of BH saddles) is possible. In fact, it is possible. \cite{Miyashitanew} (Fig. \ref{2})

For the metric and gauge field (\ref{BHf}), (\ref{BHA}), we still have regular geometries when $r_{H}>r_{b}$. These are ``bag of gold(BG)'' saddles. The role of these saddles in pure gravity is discussed in \cite{Miyashita, DraperFarkas, BanihashemiJacobson} and their importance for the 4-dimensional Einstein-Maxwell system was recently shown in \cite{Miyashitanew}. To distinguish them from  BHs, let's use $r_{G}$ for the horizon radius of BG saddles.  The expression of boundary conditions, thermodynamical quantities, and free energy are all the same except for the following two points \cite{Miyashitanew}; (i) Since $r$ is increasing toward the bolt, $f^{\p}(r_{G})$ is negative. So there is a minus sign in the relation between $\beta$ and $f^{\p}(r_{G})$ (\ref{bdyBGbeta}). (ii) Since the $r$ component of the normal vector of the boundary is negative for the same reason as before, $Q$ is not $\ti{Q}$, but $Q=-\ti{Q}$ (\ref{BGQ}). Summarize, 
%%%%%%%%%%%%%%%%%%
                                                 %
\iffigure
\begin{figure}
\begin{center}
	\includegraphics[width=8.cm]{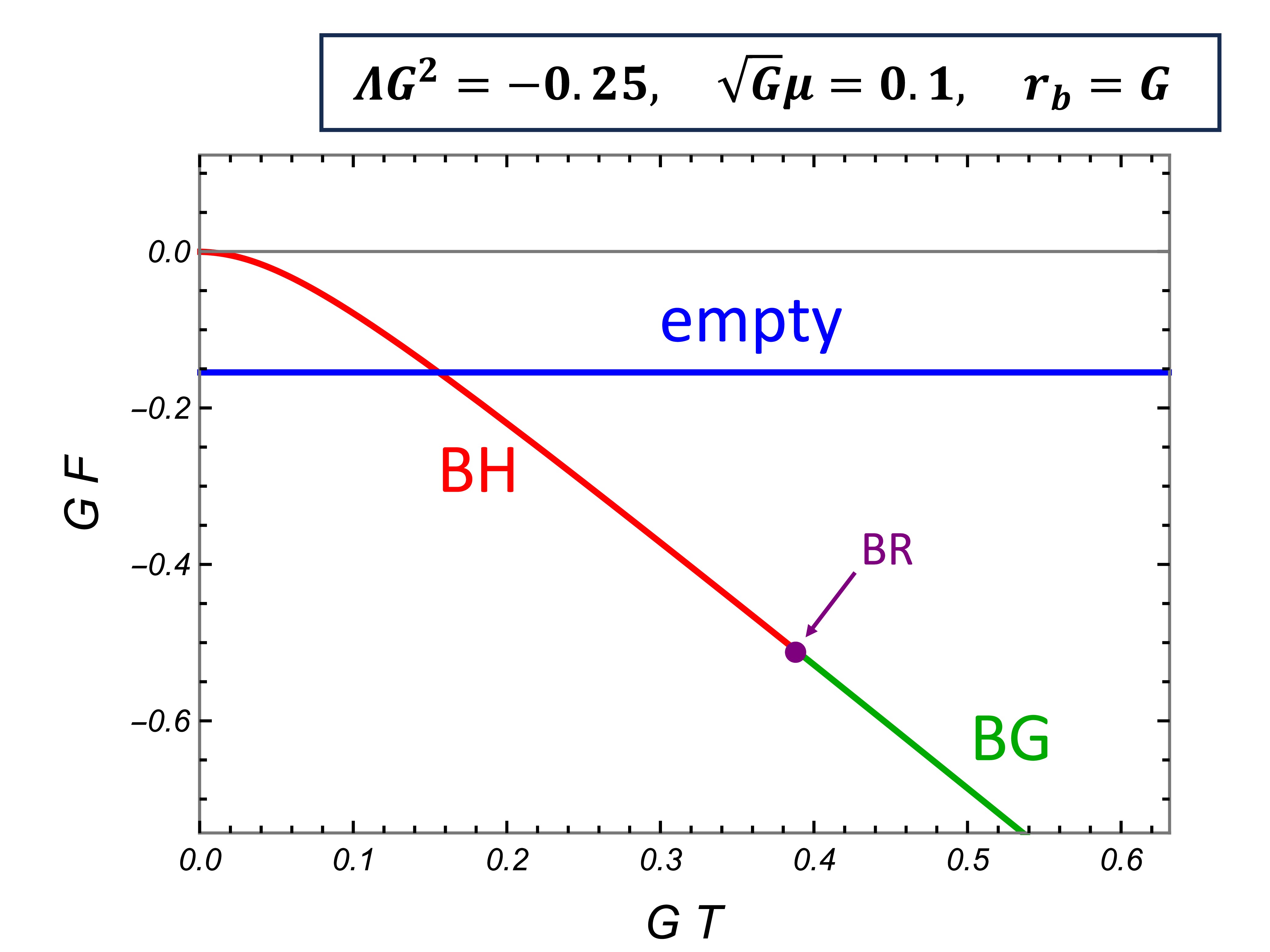} 
		\includegraphics[width=8.cm]{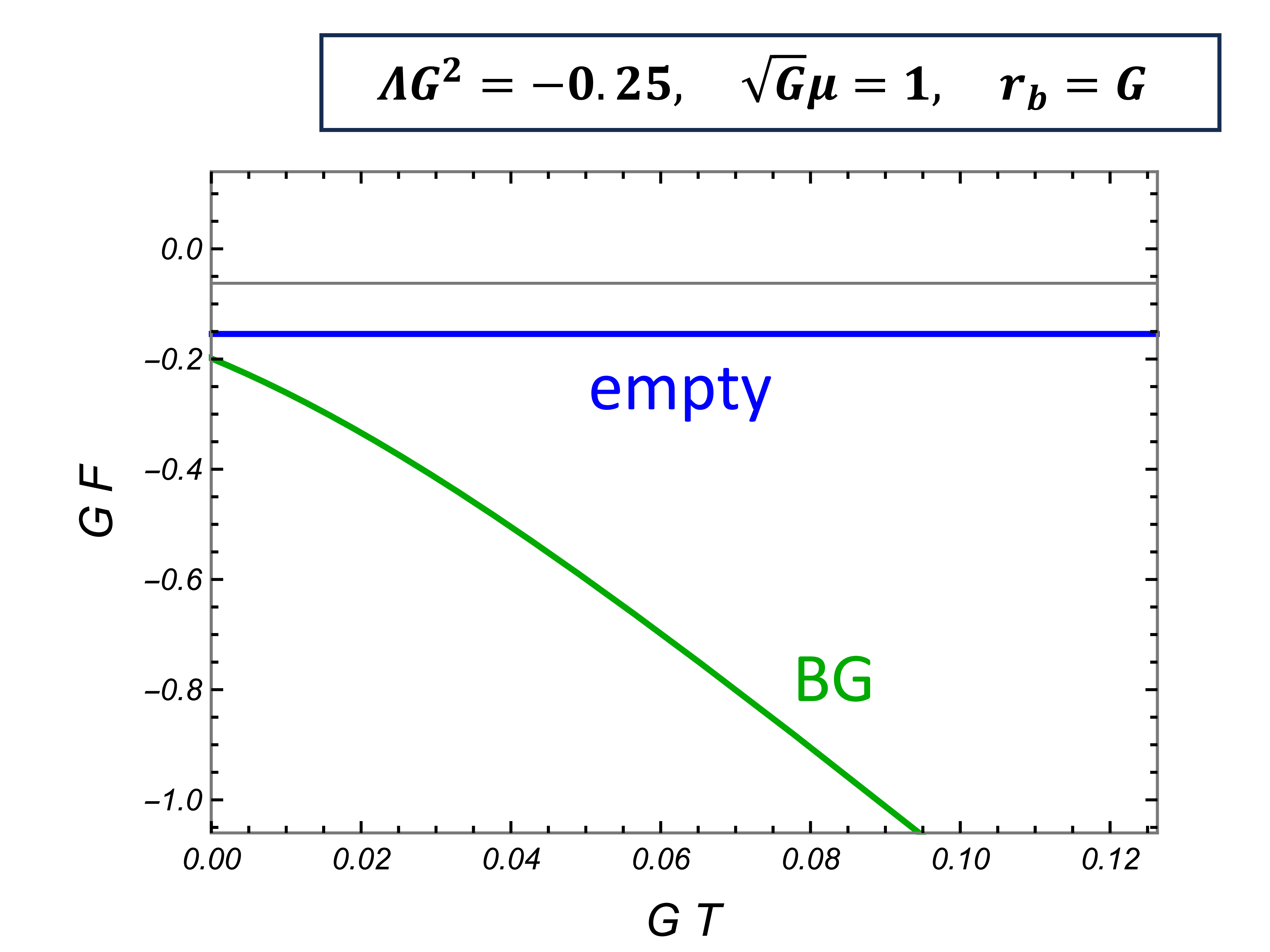} 
	\caption{Behavior of the free energies. The terminal point of the BH branch corresponds to the BR saddle. Above the terminal temperature, there are BG phases which are thermodynamically stable. For low $\mu$, there is a Hawking-Page phase transition between the BH/BG phases as shown in the left figure. For high $\mu$, the Hawking-Page phase transition does not occur and the system is in the BH/BG phase for all $T$ as shown in the right figure. The left figure is the correct version of Fig. \ref{1}.  }
\label{3}
\end{center}
\end{figure}
\fi
                                                 %
%%%%%%%%%%%%%%%%%%
\bea
{\rm \underline{boundary ~ condition}} \hspace{11.2cm} \notag \\
{\rm BG ~ saddle:} ~~ \frac{-4 \pi}{f^{\p}(r_{G})}\sqrt{f(r_{b})}= \beta, ~~ \frac{2\ti{Q}}{\sqrt{f(r_{b})}} \log \frac{r_{b}}{r_{G}} =\mu \label{bdyBGbeta} \hspace{2.95cm} 
\ena
\bea
{\rm \underline{free ~ energy}} \hspace{12.7cm} \notag \\
{\rm BG ~ saddle:} ~~ F=  \frac{\sqrt{f(r_{b})}}{4G} + \frac{\sqrt{|\Lambda|}}{4G}r_{b} -\frac{\Lambda r_{G}^2}{4G \sqrt{f(r_{b})}} - \frac{\ti{Q}^2}{\sqrt{f(r_{b})}} +\frac{2\ti{Q}^2}{\sqrt{f(r_{b})}} \log \frac{r_{b}}{r_{G}} \hspace{0.3cm}
\ena
\bea
{\rm \underline{thermodynamical ~ quantity}} \hspace{10cm} \notag \\
{\rm BG ~ saddle:} ~~ E= \frac{\sqrt{f(r_{b})}}{4G} + \frac{\sqrt{|\Lambda|}}{4G} r_{b}  \hspace{7.2cm} \\
S= \frac{\pi r_{G}}{2G} \hspace{9.5cm} \\
Q=-\ti{Q} \hspace{9.65cm} \label{BGQ}
\ena
The parameter range of $r_{G}$ is $r_{G}\in \left[r_{b}, r_{b}e^{2G\mu^2} \right) $ and the $r_{G}\to r_{b}e^{2G\mu^2}$ limit corresponds to $T\to \infty$. Therefore, the BG branch is connected to the BH branch and is always present and is dominant above the BR temperature. The examples  of the free energy behavior are shown in Fig. \ref{3}. 
The behavior of the phase diagram depends on the value $-\Lambda r_{b}^2$ and are classified into three cases; the case of $0<-\Lambda r_{b}^2 <1$, $-\Lambda r_{b}^2= -1$, and $1<-\Lambda r_{b}^2$. They are shown in Fig. \ref{4}. 
%%%%%%%%%%%%%%%%%%
                                                 %
\iffigure
\begin{figure}[h]
\begin{center}
	\includegraphics[width=5.3cm]{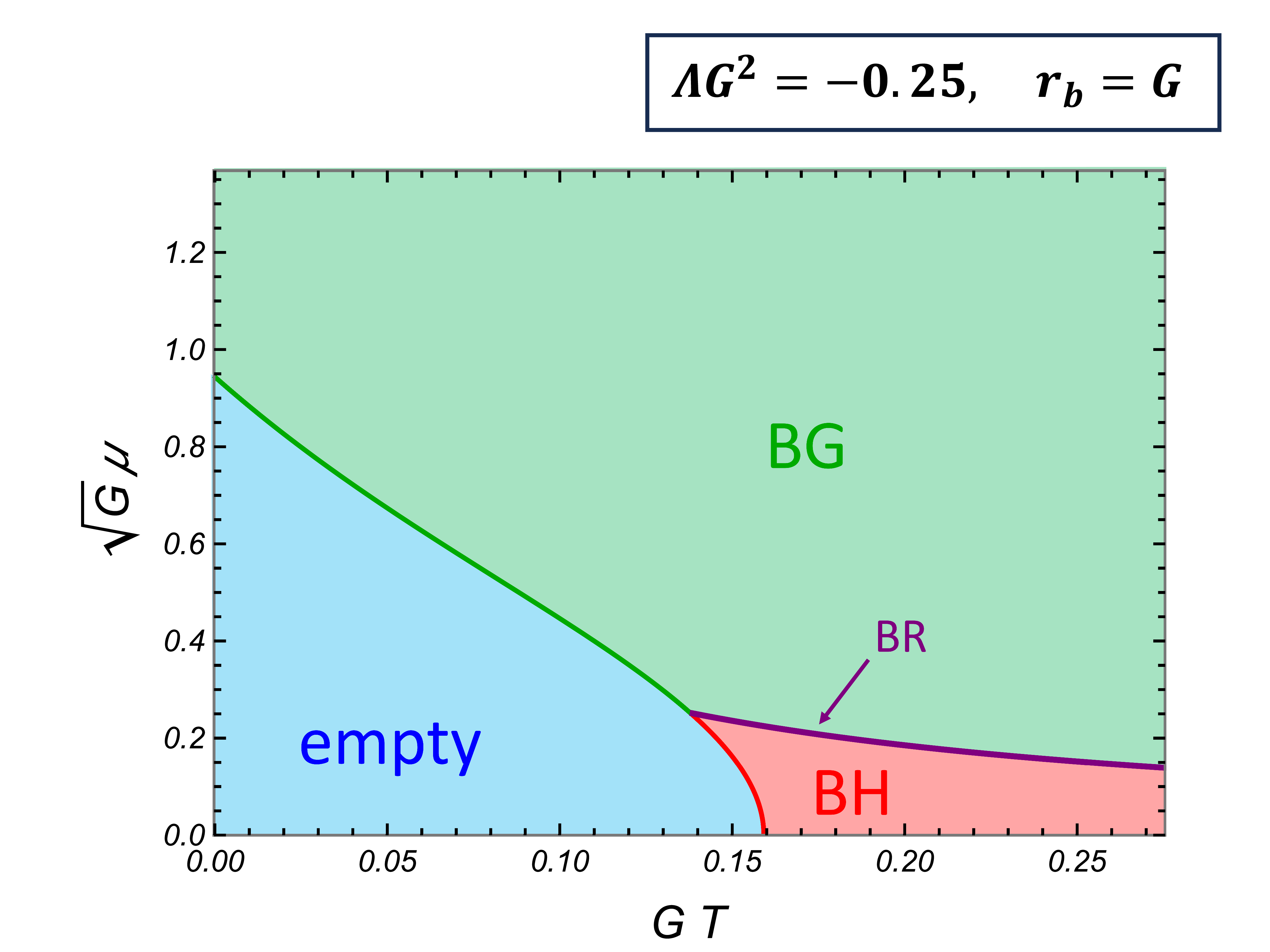} 
	\includegraphics[width=5.3cm]{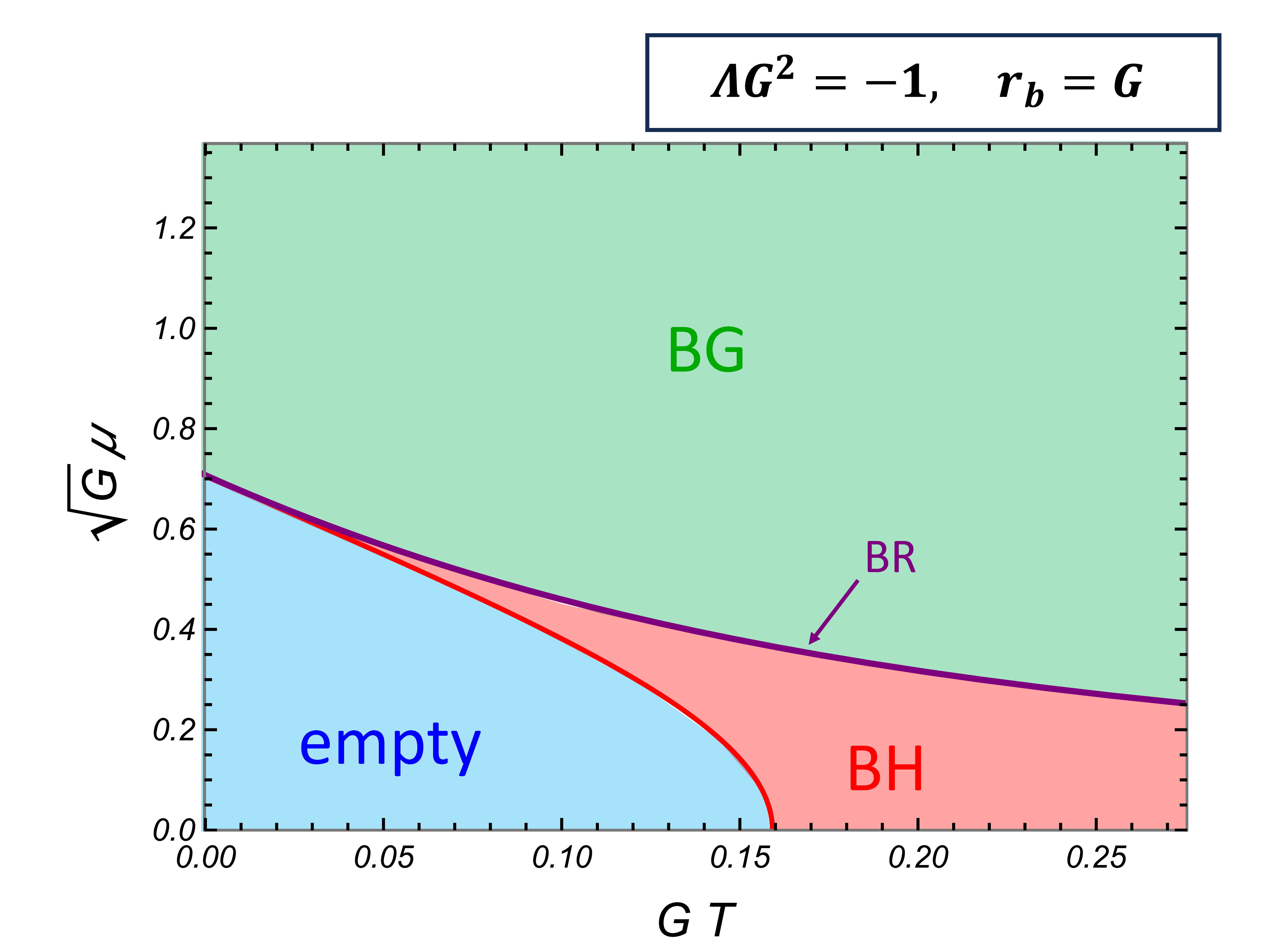} 
	\includegraphics[width=5.3cm]{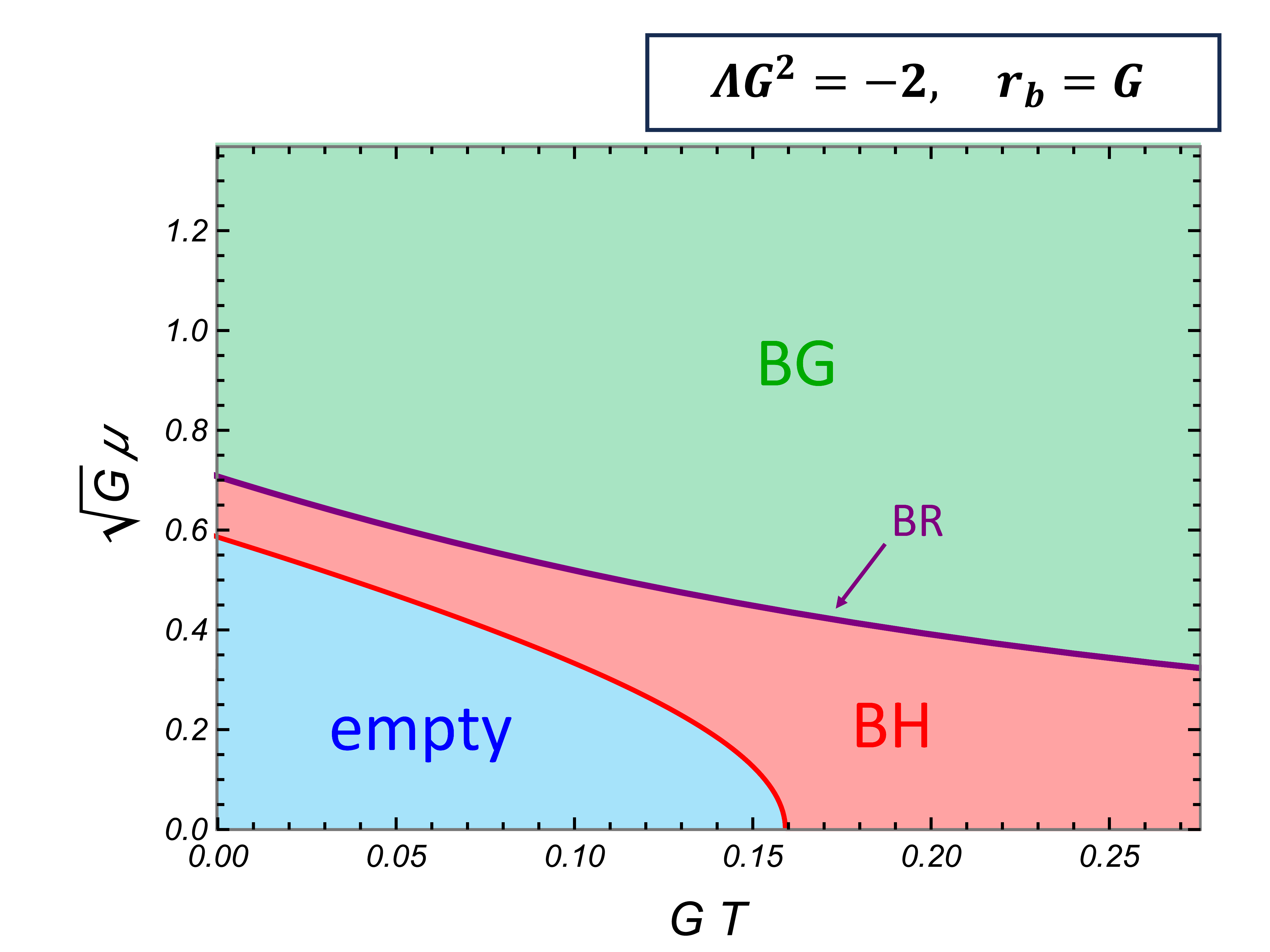} 
	\caption{Examples of phase diagrams. The behavior of the phase diagram depends on the value of $-\Lambda r_{b}^2$ and is classified into three cases; the case of $0<-\Lambda r_{b}^2 <1$ (Left), $-\Lambda r_{b}^2= -1$ (Middle), and $1<-\Lambda r_{b}^2$ (Right). The left figure is the correct version of Fig. \ref{1}. }
\label{4}
\end{center}
\end{figure}
\fi
                                                 %
%%%%%%%%%%%%%%%%%%

%======================================%
%<<<<<<<<<<<< SECTION III  >>>>>>>>>>>>>>%
%======================================%
%%%%%%%%%%%%%%%%%%%%%%%%%%%%%%%%%%%%%%%%%%%
%%%%%%%%%%%%%%%%%%%%%%%%%%%%%%%%%%%%%%%%%%%
%%%%%%%%%%%%%%%%%%%%%%%%%%%%%%%%%%%%%%%%%%%
%%%%%%%%%%%%%%%%%%%%%%%%%%%%%%%%%%%%%%%%%%%
\section{Difference between 3-dim. and 4-dim. ($\Lambda < 0$)}
%%%%%%%%%%%%%%%%%%%%%%%%%%%%%%%%%%%%%%%%%%%%%%%%%%%
%%%%%%%%%%%%%%%%%%%%%%%%%%%%%%%%%%%%%%%%%%%%%%%%%%%
%%%%%%%%%%%%%%%%%%%%%%%%%%%%%%%%%%%%%%%%%%%%%%%%%%%
There are many qualitative and quantitative differences between the Einstein-Maxwell system in 3-dimensions and that in 4-dimensions.
%%%%%%%%%%% footnote %%%%%%%%%%%%%%
\footnote{
For example, 
\begin{itemize}
\item There is no upper bound on $\mu$ for the BH and BG saddles in 3-dim., but $\sqrt{G}\mu=1$ is the upper bound in 4-dim. 
\item In both cases, there exists the maximum horizon radius for fixed $\mu$. $r_{G, max} = r_{b}e^{2G\mu^2}$ for 3-dim. and $r_{G, max}= r_{b} \frac{1}{1-G\mu^2}$ for 4-dim.
\item In 4-dim., there can exist another small BH phase in the low temperature region, separated from the main BH phase, when $-\Lambda r_{b}^2$ is in a certain range. Compare Fig \ref{4} and Fig. 11 in \cite{Miyashitanew}.
\item When $\mu =0$, the transition temperature is always given by $T= \frac{1}{2\pi r_{b}}$ in 3-dim. and does not depend on $\Lambda$, as shown in \cite{HuangTao}. In 4-dim. it also depends on $\Lambda$.   
\end{itemize}
}
%%%%%%%%%%%%%%%%%%%%%%%%%%%%%%%%%
What I want to emphasize here the most is the behavior of the BH phase (and the BG phase) in the phase diagram versus the value of $ \Lambda $. In 4-dimensions, when we take $\Lambda \to 0$ limit or $\Lambda=0$, the phase diagram is still similar to the left figure in Fig. \ref{4}, i.e. there is a BH phase \cite{Miyashitanew}. However, in 3-dimensions, if we decrease the value of $-\Lambda r_{b}^2$, the BH phase shrinks. This can be seen from the fact that the boundary curve between the BH phase and the BG phase is given by (\ref{TBR}) and the curve becomes close to the $T$ axis. This indicates that the BH phase disappears when $\Lambda=0$. The disappearance of the BH phase is not surprising, since, under some suitable conditions, including the dominant energy condition, it is proved that there are no BHs in 3-dimensions when $\Lambda = 0$ \cite{Ida}. Even for the system with a box boundary, this statement still be true \cite{KrishnanShekharSubramanian}. What may be surprising is the existence of the BG phase even for the $\Lambda\to0$ limit. This seems to imply that even for $\Lambda = 0$, the BG phase exists, possibly as a thermodynamically stable state. However, I will show in the next section that BG saddles do exist, but they cause thermodynamic instability.

%======================================%
%<<<<<<<<<<<< SECTION IV  >>>>>>>>>>>>>>%
%======================================%
%%%%%%%%%%%%%%%%%%%%%%%%%%%%%%%%%%%%%%%%%%%
%%%%%%%%%%%%%%%%%%%%%%%%%%%%%%%%%%%%%%%%%%%
%%%%%%%%%%%%%%%%%%%%%%%%%%%%%%%%%%%%%%%%%%%
%%%%%%%%%%%%%%%%%%%%%%%%%%%%%%%%%%%%%%%%%%%
\section{$\Lambda = 0$ case}
%%%%%%%%%%%%%%%%%%%%%%%%%%%%%%%%%%%%%%%%%%%%%%%%%%%
%%%%%%%%%%%%%%%%%%%%%%%%%%%%%%%%%%%%%%%%%%%%%%%%%%%
%%%%%%%%%%%%%%%%%%%%%%%%%%%%%%%%%%%%%%%%%%%%%%%%%%%
In this case, there exist two types of saddles, the one is the empty saddle, 
\bea
ds^2 = g_{\mu\nu}dx^{\mu}dx^{\nu} = dt^2 + dr^2 + r^2 d\phi^2
\ena
And the other type of saddles can be obtained by setting $\Lambda =0$ in (\ref{BHf}), (and replacing the parameter $r_{H}$ with $r_{G}$,)
\bea
f(r) =  -  8GQ^2 \log \frac{r}{r_{G}}  \\
A_{t}(r) = 2 i Q \log \frac{r}{r_{G}} 
\ena
Since for the coordinate range $r\in[r_{b}, r_{G})$, $f(r)>0$ and $f(r_{G})=0$, this is the BG saddle, which is not singular. And here I also set $\ti{Q}=-Q$ since these are BGs. The boundary conditions are similar to those in subsection 2.2. From the boundary conditions, we get the following expression;
\bea
\frac{1}{2G} \log \frac{r_{G}}{r_{b}} = \mu^2 \label{flatmu} \\
T= \frac{Q}{2\pi r_{b}} \frac{1}{\mu e^{2G\mu^2}}
\ena
At this point, if we admit a general fact that the Bekenstein-Hawking formula $S = \frac{\pi r_{G}}{2G}$ is true,
%%%%%%%%%%%% footnote %%%%%%%%%%%%%%%
\footnote{
Strictly speaking, we do not have to use the Bekenstein-Hawking formula here and we can confirm this relation $S= \frac{\pi r_{G}}{2G}$ by calculating the free energy and using thermodynamic relations, as I will derive in (\ref{freeBGentropy}).
}
%%%%%%%%%%%%%%%%%%%%%%%%%%%%%%%%%%%
 we can easily see that the entropy does not change no matter how we increase or decrease the temperature for fixed $\mu$ (i.e., from (\ref{flatmu}) and the Bekenstein-Hawking formula, the entropy $S(T,\mu)$ can be written as $S= \frac{\pi r_{G}}{2G}=\frac{\pi r_{b}}{2G} e^{2G\mu^2} $ which does not depend on $T$). This implies that the heat capacity is zero. Therefore, if these saddles give dominant contribution, it leads to thermodynamical instability of the system. 
%%%%%%%%% footnote %%%%%%%%%%%%%%%
\footnote{
Precisely, it is neither stable nor unstable, it is marginal. However, if it is not stable, the system may not reach thermal equilibrium. It is in this sense that I refer to this situation as unstable.
}
%%%%%%%%%%%%%%%%%%%%%%%%%%%%%%%%
Explicitly, free energy and thermodynamical quantities are given by
\bea
{\rm \underline{free ~ energy}} \hspace{12.7cm} \notag \\
{\rm empty ~ saddle:} ~~ F= \frac{-1}{4G} \hspace{8.95cm} \\
{\rm BG ~ saddle:} ~~ F= - \frac{Q^2}{\sqrt{f(r_{b})}} -\frac{2Q^2}{\sqrt{f(r_{b})}} \log \frac{r_{b}}{r_{G}}  \hspace{5cm} \notag \\
= -\frac{ \pi r_{b} T}{2G} \left(1+ 4G \mu^2 \right) e^{2G\mu^2} \label{freeflat} \hspace{5.4cm}
\ena
\bea
{\rm \underline{thermodynamical ~ quantity}} \hspace{10cm} \notag \\
{\rm empty ~ saddle:} ~~ E= \frac{-1}{4G} \hspace{8.95cm} \\
S= 0 \hspace{9.35cm} \\
Q= 0 \hspace{9.35cm} \\
~\notag \\
{\rm BG ~ saddle:} ~~ E= 0 \hspace{9.4cm} \\
S= \frac{\pi r_{G}}{2G} = \frac{\pi r_{b}}{2G} e^{2G\mu^2} \label{freeBGentropy} \hspace{6.8cm} \\
Q= 2\pi r_{b}T \mu e^{2G\mu^2} \hspace{7.45cm}
\ena
Since, in this case, the BG free energy (\ref{freeflat}) is a linear function of $T$, the ``transition temperature'' can be easily found as
\bea
T_{tr} = \frac{1}{  2 \pi  r_{b} \left( 1 + 4 G  \mu^2 \right) e^{2G\mu^2} }
\ena
The ``phase diagram'' is shown in Fig. \ref{5}. 
%%%%%%%%%%%%%%%%%%
                                                 %
\iffigure
\begin{figure}
\begin{center}
	\includegraphics[width=7cm]{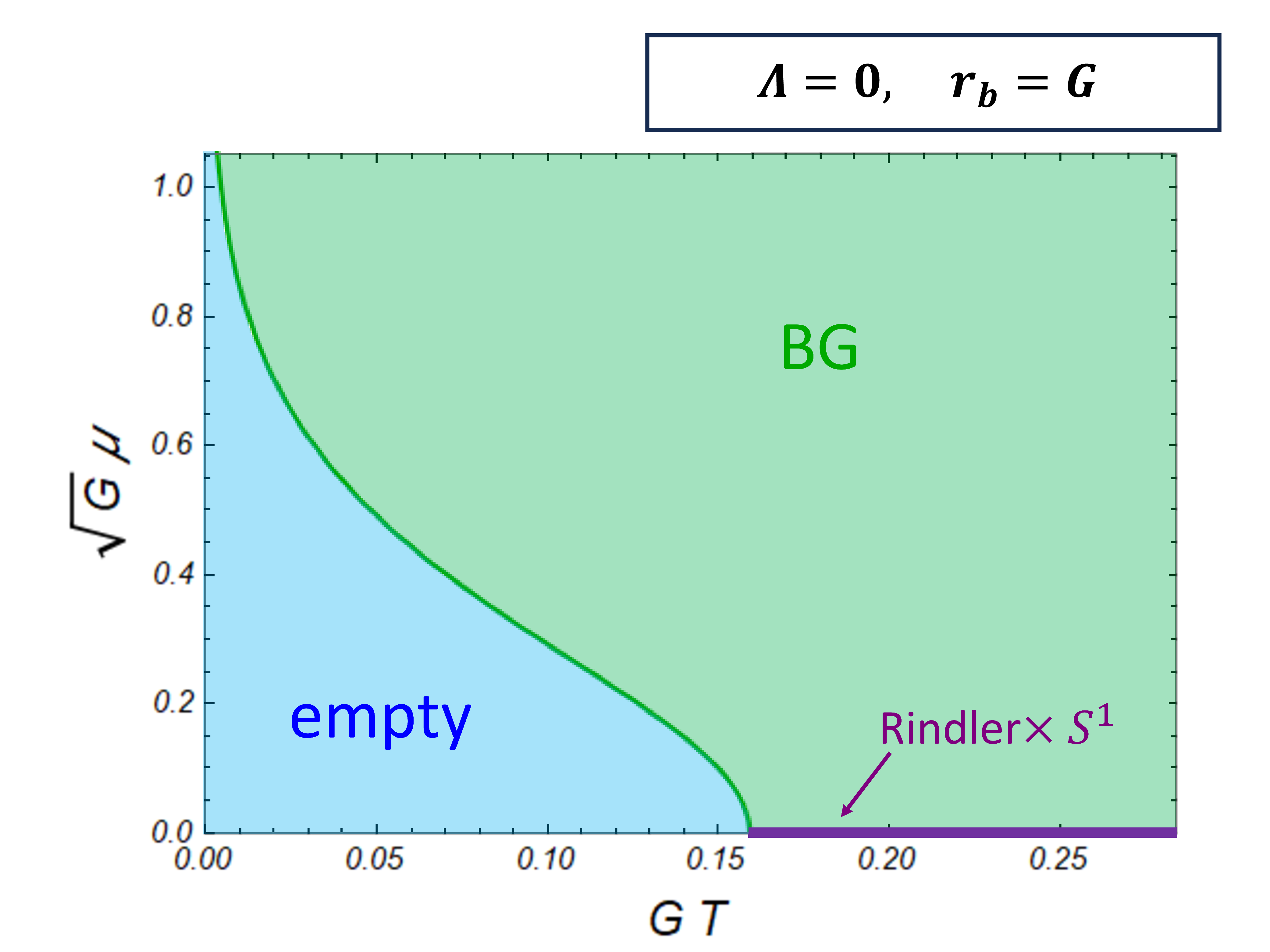} 
	\caption{``Phase diagram'' of the case where $\Lambda=0$. Since the BG ``phase'' is not thermodynamically stable and the corresponding saddles give the dominant contribution, the system itself is not thermodynamically stable either. }
\label{5}
\end{center}
\end{figure}
\fi
                                                 %
%%%%%%%%%%%%%%%%%%
Finally, let's discuss what happens when we turn off $\mu$. From (\ref{flatmu}), we see that $\mu \to 0$ limit corresponds to $r_{G} \to r_{b}$ limit. Again, this limit does not lead to the merging of horizon and boundary but leads to a regular geometry, which is the direct product of the 2-dimensional Rindler space and $S^{1}$. By defining a small parameter $\varepsilon$ by $\varepsilon=G\mu^2$, the first derivative of $f$ at $r=r_{G}$ is 
\bea
f'(r_{G}) = - \frac{8G Q^2}{r_{G}} =- 32 \pi^2 r_{b} T^2 \varepsilon + O(\varepsilon^2)
\ena
Since $r_{G}=r_{b}e^{2\varepsilon} = r_{b} + 2r_{b}\varepsilon + O(\varepsilon^2)$, let's define new coordinates by $(r_{G}-r)= 2r_{b}\varepsilon \ti{r}, ~ t=\frac{\ti{t}}{2r_{b}\varepsilon}$. Then, the ranges of these new coordinates will be $\ti{r}\in [0,1], ~ \ti{t} \in \left[ 0, \frac{1}{4\pi T^2} \right]$ in the $\varepsilon\to 0$ limit. Then, in the $\varepsilon\to 0$ limit, the metric and gauge field become
%%%%%%%%%%%%%%%% footnote %%%%%%%%%%%%%
\footnote{
An alternative way is to take a double scaling limit of the BR geometry (\ref{BRmetric}):
\beann
4\pi T = \sqrt{\frac{-\Lambda}{G\mu^2}} {\rm : fixed,} ~~ ~~ \Lambda \to0, ~ G\mu^2 \to 0
\enann
}
%%%%%%%%%%%%%%%%%%%%%%%%%%%%%%%%%%%%%
\bea
ds^2= \left[ 64\pi r_{b}^2 T^2 \varepsilon^2 \ti{r} + O(\varepsilon^3) \right] \frac{d\ti{t}^2}{ 4 r_{b}^2 \varepsilon^2} + \frac{4r_{b}^2 \varepsilon^2 }{ \left[ 64\pi r_{b}^2 T^2 \varepsilon^2 \ti{r} + O(\varepsilon^3) \right] } d\ti{r}^2 + (r_{b}+O(\varepsilon))^2 d\phi^2 \notag \\
\to 16\pi^2 T^2 \ti{r} d\ti{t}^2 + \frac{1}{16\pi^2 T^2 \ti{r}}d\ti{r}^2 + r_{b}^2 d\phi^2 \hspace{6.5cm}  \\
\BS{A} \to 0 \hspace{12.15cm}
\ena
If we further define the coordinates $\rho$ and $\tau$ by $\ti{r} = 4\pi^2 T^2 \rho^2, ~ \tau=8 \pi^2 T^2 \ti{t}$, the metric takes the standard Rindler form
\bea
ds^2 = \rho^2 d\tau^2 + d\rho^2 + r_{b}^2 d\phi^2 \\
\tau\in [0, 2\pi], ~ \rho \in [0, 2\pi T]
\ena
The free energy of the $Rindler \times S^{1}$ saddles is given by simply setting $\mu=0$ in the free energy of the BG saddles (\ref{freeflat}), and we can easily check that it will be dominant above the temperature $T= \frac{1}{2\pi r_{b}}$. (Fig. \ref{5})

At first glance, the phase structures of the Einstein-Maxwell system in 3-dimensions and 4-dimensions are similar when $-\Lambda r_{b}^2$ is sufficiently large. There are three phases, the empty phase, the BH phase, and the BG phase (Fig. \ref{4}). However, when it is close to zero, differences appear; For 4-dimensions, the BH phase exists when $\Lambda \to 0$ or $\Lambda =0$, and the system is thermodynamically stable. For 3-dimensions, however, it shrinks in the $\Lambda \to 0$ limit and ceases to exist at $\Lambda = 0$. So in 3-dimensions, when a negative $\Lambda$ is switched off, the boxed systems no longer have the saddles with the smaller horizon (i.e. BH saddles), as previously proved  \cite{Ida, KrishnanShekharSubramanian}, but still have the ones with the same size or larger horizon (i.e. BR saddles and BG saddles). But these saddles make the system thermodynamically unstable. 
%%%%%%%%%%%%%%%% footnote %%%%%%%%%%%%%
\footnote{
When $\Lambda = 0$ and $\mu>0$, the BG saddle is not completely unstable, but it is a kind of marginal point, in the sense that its heat capacity is not negative, but zero. It would be interesting to study the existence of hairy extensions of BG saddles and their thermodynamical stability when $\Lambda=0$.
}
%%%%%%%%%%%%%%%%%%%%%%%%%%%%%%%%%%%%%
If we also turn off $\mu$, the system does not allow the saddle with the larger horizon (BG saddle), but still allows the ones with the same size horizon (BR saddle, or $Rindler \times S^{1}$ saddle), which again makes the system thermodynamically unstable. 
%%%%%%%%%%%%%%%% footnote %%%%%%%%%%%%%
\footnote{
For the $\Lambda > 0$ case, there are the empty phase and the BG phase. The form of the metric and gauge field are the same as in the $\Lambda<0$ case. The coordinate range of $r$ can be classified by the value of $\mu$; $r\in [r_{b}, \infty) $ for $2G\mu^2 \leq 1$, and $r\in \left[r_{b}e^{2G\mu^2}, \infty\right) $ for $1<2G\mu^2$. But in both cases, the BG saddles are thermodynamically unstable and so is the system. This situation is the same as in the 4-dimensional case \cite{Miyashitanew}. 
}
%%%%%%%%%%%%%%%%%%%%%%%%%%%%%%%%%%%%%

This thermodynamical instability offers two options for how we think about the thermodynamics of 3-dimensional gravity with $\Lambda=0$. One is that the zero-loop approximation or restricting the simple topology sector is not sufficient for 3-dimensions with $\Lambda=0$. The other is that 3-dimensional gravity is really thermodynamically unstable when $\Lambda=0$.

\section*{Acknowledgement}
This work is supported in part by the National Science and
Technology Council (No. 111-2112-M-259-016-MY3).

%%%%%%%%%%%%%%%%%%%%%%%%%%%%%%%%%%%%%%
\appendix
\section{Off-shell geometries for M states}
This appendix is purely an appendix and has no direct relation to the result in this paper. In \cite{HuangTao}, they claimed that the $r_{H}\to r_{b}$ limit of the BH leads to the merging of the horizon and the boundary. Even for such non-geometrical states, they claimed that there are thermodynamical meaning and assigned the free energy of the form 
\bea
F_{M state}(T, \mu) = \frac{\sqrt{-\Lambda}}{4G}r_{b} - T \frac{\pi r_{b}}{2G} - \mu \frac{r_{b}}{2}\sqrt{\frac{-\Lambda}{G}} \label{FMstate}
\ena  

But in subsection 2.2, I showed that the limit corresponds to the BR geometry, which is a regular geometry and it does not have arbitrary temperature and chemical potential, but they have a relationship (\ref{TBR}). Although the geometric picture of the M states was wrong, the free energy of the M states at the BR point was correct.
\bea
F_{M state} (T_{BR}(\mu), \mu) = F_{BR}(\mu)
\ena
In other words, at this BR point, the M state has a geometric picture, which is the on-shell BR geometry. So one might expect that other M states deformed from the M state at the BR point might have a geometric picture. Here, I show that BR geometries with a conical singularity 
\bea
ds^2= \left[ \frac{-\Lambda}{G \mu^2} (1-2G \mu^2) \ti{r} - 2\Lambda \ti{r}^2 \right] d\ti{t} + \frac{1}{\left[ \frac{-\Lambda}{G \mu^2} (1-2G \mu^2) \ti{r} - 2\Lambda \ti{r}^2 \right]} d\ti{r}^2 + r_{b}^2 d\phi^2 \label{BRsingular} \\
\BS{A}=  -i \sqrt{\frac{-\Lambda }{G}} \ti{r} d\ti{t} \hspace{9.75cm} \\
\ti{r}\in[0, 1], ~~~~~~~ \ti{t}\in \left[0, \beta \mu \sqrt{\frac{G}{-\Lambda}} \right] \hspace{4.5cm} \notag 
\ena
can be a family of off-shell geometries whose free energy is given by (\ref{FMstate}). 
%%%%%%%%%%%%%% footnote %%%%%%%%%%%%%%%
\footnote{
Of course, there are infinitely many families of off-shell geometries satisfying (\ref{FMstate}). This is just a simple example, and probably the simplest.
}
%%%%%%%%%%%%%%%%%%%%%%%%%%%%%%%%%%%%%%
The parameters of this geometry are $r_{b}, \mu$, and $\beta$, where $\beta$ is the circumference of the Euclidean time circle at the boundary. When $\DS \beta \neq \beta_{BR} = \frac{4\pi}{\sqrt{-\Lambda}} \frac{\sqrt{G}\mu}{ 1-2 G \mu^2 } $, this geometry has a conical singularity at $\ti{r}=0$, thus it is off-shell. The action functional for geometries with a conical singularity is given by \cite{FursaevSolodukhin}
\bea
I^{E}[\BS{g}, \BS{A}]= \frac{-1}{16\pi G} \int_{\MC{M}\setminus\Sigma} d^3 x \sqrt{g} (\MC{R}- 2\Lambda) + \frac{1}{16 \pi}\int_{\MC{M}\setminus\Sigma} d^3 x \sqrt{g} F_{\mu\nu} F^{\mu\nu} \hspace{4.cm} \notag \\
+ \frac{-1}{8\pi G} \int_{\pd \MC{M}} d^2 y \sqrt{\gamma} \Theta + \frac{1}{8\pi G} \int_{\pd \MC{M}} d^2 y \sqrt{\gamma}  \sqrt{|\Lambda|} + \frac{-1}{8\pi G} \int_{\Sigma} d\phi \sqrt{\sigma} (2\pi -\theta(\phi))
\ena
The last term is for conical singularities and $\Sigma$ represents the codimension-two surface of a conical singularity and $2\pi-\theta(\phi)$ represents the (position dependent) conical deficit on the surface. The bulk and boundary terms are the same as for a regular BR, i.e. they are given by $\beta$ times $F_{BR}$ (\ref{BRfree}). The additional contribution comes from the last term. Substituting (\ref{BRsingular}) for it gives
\bea
\frac{-1}{8\pi G} \int_{\Sigma} d\phi \sqrt{\sigma} (2\pi -\theta(\phi)) = \frac{-r_{b}}{4G} \left( 2\pi + \frac{\Lambda}{2G\mu^2} (1-2G\mu^2) \beta \mu \sqrt{\frac{G}{-\Lambda}}  \right) \notag\\
= - \frac{\pi r_{b}}{2G} +  \frac{r_{b}}{8G\mu} \sqrt{\frac{-\Lambda}{G}} \beta - \frac{r_{b}}{4}\sqrt{\frac{-\Lambda}{G}}  \beta \mu \hspace{0.8cm} \label{FS}
\ena
Therefore, the ``off-shell'' free energy of these singular BR geometries is given by
\bea
F_{singular BR}= F_{BR} + T \times (\ref{FS}) \hspace{8.75cm} \notag \\
= \frac{\sqrt{-\Lambda}}{4G}r_{b} - \frac{r_{b}}{8G\mu}\sqrt{\frac{-\Lambda}{G}} -\frac{r_{b}\mu}{4} \sqrt{ \frac{-\Lambda}{G} }  - \frac{\pi r_{b}}{2G}T + \frac{r_{b}}{8G\mu} \sqrt{\frac{-\Lambda}{G}}  -  \frac{r_{b}}{4} \sqrt{\frac{-\Lambda}{G}} \mu \notag \\
= \frac{\sqrt{-\Lambda}}{4G}r_{b} - \frac{\pi r_{b}}{2G}T   -  \frac{r_{b}}{2} \sqrt{\frac{-\Lambda}{G}} \mu \hspace{6.75cm}
\ena
which is equivalent to the free energy of the M state (\ref{FMstate}).

%%%%%%%%%%%%%%%%%%%%%%%%%%%%%%%%%%%%%%%%%%%%%%%%%%%
%%%%%%%%%%%%%%%%%%%%%%%%%%%%%%%%%%%%%%%%%%%%%%%%%%%
%%%%%%%%%%%%%%%%%%%%%%%%%%%%%%%%%%%%%%%%
%%%%%%%%%%%%%%%%%%%%%%%%%%%%%%%%%%%%%%%%%%%%%%%%%%%
%%%%%%%%%%%%%%%%%%%%%%%%%%%%%%%%%%%%%%%%%%%%%%%%%%%
%%%%%%%%%%%%%%%%%%%%%%%%%%%%%%%%%%%%%%%%
\appendix

\end{document}